\documentclass[preprint, 12pt]{elsarticle}
\usepackage{latexsym}
\usepackage{amsmath}
\usepackage{amssymb}
\usepackage{amsthm}
\usepackage{amscd}
\usepackage[mathscr]{eucal}

\usepackage{graphicx}
\usepackage{subfigure}

\theoremstyle{plain}

\begin{document}

\begin{frontmatter}
\title{Cosmological models with fluid matter\\ undergoing velocity diffusion}

\author{Simone Calogero}
\ead{calogero@ugr.es}
\address{Departamento de Matem\'atica Aplicada\\
Facultad de Ciencias, Universidad de Granada
18071 Granada, Spain}

\begin{abstract}
A new type of fluid matter model in general relativity is introduced, in which the fluid particles are subject to velocity diffusion without friction. In order to compensate for the energy gained by the fluid particles due to diffusion, a cosmological scalar field term is added to the left hand side of the Einstein equations. This hypothesis  promotes diffusion to a new mechanism for accelerated expansion in cosmology. 
It is shown that diffusion alters not only quantitatively, but also qualitatively the global dynamical properties of the standard cosmological models.
\end{abstract}
\end{frontmatter}

\section{Introduction}
The cause of several physical and biological processes can be attributed to diffusion.  
Notable examples are heat conduction, Brownian motion and the transport of materials within cells. 
At the microscopic level, diffusion is due to random collisions between the particles of the system under study and those of a background substance. At the macroscopic scale, in which random effects are averaged, diffusion is described by an effective, deterministic theory. The macroscopic physical quantities are solutions of partial differential equations, such as the heat equation or the Fokker-Planck equation. The applications of such theories go well beyond the original physical motivations and include fields as different as image processing, economics and social sciences.

The purpose of this paper is to propose and explore the idea that diffusion may also play a fundamental role in the large scale dynamics of the matter in the universe. The standard assumption in cosmology is that the galaxies (in which most of ordinary matter is supposed to be concentrated) can be represented as the particles of a fluid. Here it will be further  assumed that the fluid particles undergo velocity diffusion. The model will be developed within Einstein theory of general relativity with a cosmological scalar field, the latter playing the role of the background medium in which diffusion takes place.
   
\section{The model}
The general relativistic theory of diffusion processes has received considerable attention in recent years, see for instance~\cite{calogero, frajan, haba, herr}. As shown in~\cite{calogero},  the energy-momentum tensor $T^{\mu\nu}$ and the current density $J^\mu$ of matter undergoing velocity diffusion satisfy the equations $\nabla_\mu J^\mu=0$ and
\begin{equation}\label{tjeq}
\nabla_\mu T^{\mu\nu}=3\sigma J^\nu.
\end{equation}
Here $\nabla_\mu$ denotes the Levi Civita covariant differentiation on a spacetime $(M,g)$, and $\sigma$ is a positive dimensional constant. In ref.~\cite{calogero} the matter field is given by the one-particle kinetic distribution function $f$ in phase space, and the tensor fields $T^{\mu\nu}$, $J^\mu$ are defined by suitable integrals of $f$ in the momentum variable. The relation~\eqref{tjeq} is a consequence of the evolution equation satisfied by $f$, which is a degenerate Fokker-Planck equation without friction on the tangent bundle of spacetime. Hence the  model studied in~\cite{calogero} describes the kinetic motion of an ensemble of particles undergoing diffusion in the velocity variable. The right hand side of~\eqref{tjeq} is a macroscopic representation of the diffusion forces acting on the particles, and $\sigma$ is the diffusion constant, which measures the average energy transferred per
unit time from the background substance to a fluid particle.  

This paper is concerned with fluid matter models with energy-momentum tensor and current density 
\begin{equation}\label{tjfluid}
T^{\mu\nu}=\rho u^\mu u^\nu+p(g^{\mu\nu}+u^\mu u^\nu),\quad J^\mu=n u^\nu
\end{equation}
that satisfy 
\begin{align}
&\nabla_\mu T^{\mu\nu}=\sigma n u^\nu,\label{divT}\\
&\nabla_\mu (n u^\mu)=0\label{divJ}.
\end{align} 
The  scalar functions $\rho, p, n$ represent respectively the rest-frame energy density, the pressure and the number density of the fluid, while the vector field $u^\mu:u_\mu u^\mu=-1$, is the four-velocity of the fluid.
Motivated by the kinetic model introduced in~\cite{calogero}, an ideal fluid that satisfies~\eqref{tjeq} will be said to undergo (molecular) velocity diffusion and the parameter $\sigma$ will be called diffusion constant in the present context as well. Note that in the present paper the factor 3 in the right hand side of~\eqref{tjeq} is absorbed in the definition of the diffusion constant.  

Eq.~\eqref{divT} is equivalent to the following equations
\begin{align}
&\nabla_\mu(\rho u^\mu)+p\nabla_\mu u^\mu=\sigma n,\label{cont}\\
&(\rho+p)u^\mu\nabla_\mu u^\nu+u^\nu u^\mu\nabla_\mu p+\nabla^\nu p=0,\label{euler}
\end{align}
 which are obtained by projecting~\eqref{divT} into the direction of $u^\mu$ and onto the hypersurface orthogonal to $u^\mu$, respectively. Note that the Euler equation~\eqref{euler} is the same as in the diffusion-free case; only the continuity equation~\eqref{cont} is affected by the presence of diffusion. This is due to the fact that the diffusion force $\sigma n u^\mu$ acts on the direction of the matter flow.

To transform~\eqref{divJ}--\eqref{euler} into a complete system on the variables $(\rho,p,n,u^\mu)$, an equation of state must be added. 
Letting the energy per particle $e$ and the specific volume  $v$  be defined as $e=\rho/n$, and $v=1/n$,
the first law of thermodynamics states that 
\begin{equation}\label{1stlaw}
de=-pdv+T ds,
\end{equation}
where the scalar functions $T$ and $s$ denote respectively the temperature and the entropy of the fluid. Units are chosen so that Boltzmann's constant is equal to one.
An equation of state relating the thermodynamic variables may be prescribed in the form
\begin{equation}\label{eqstate}
\rho=h(n,s),
\end{equation}
for some function $h$, which for simplicity will be assumed to be continuously differentiable.
Substituting in~\eqref{1stlaw} we obtain
\begin{equation}\label{enthalpy}
\mathfrak{e}:=\frac{\rho+p}{n}=\frac{\partial h}{\partial n}, \quad T(n,s)=\frac{1}{n}\frac{\partial h}{\partial s}.
\end{equation} 
For ordinary fluids the specific enthalpy $\mathfrak{e}$ and the temperature $T$ are required to be positive, which implies that
$\rho+p>0$, $\partial h/\partial n>0$, and $\partial h/\partial s>0$.
Then by~\eqref{cont} and~\eqref{divJ} we obtain
\begin{equation}\label{enteq}
u^\mu\nabla_\mu s=\frac{\sigma}{T(n,s)}.
\end{equation} 
The equations~\eqref{cont},~\eqref{euler},~\eqref{eqstate},~\eqref{enthalpy},~\eqref{enteq} provide the desired complete system for describing the evolution of the matter field variables $(n,s,u^\mu)$. Eq.~\eqref{cont} can be replaced by eq.~\eqref{divJ}, since they are equivalent when~\eqref{enteq} holds.
Note also that the entropy of the fluid is strictly increasing along the matter flow, unless diffusion is switched off (i.e., $\sigma$ is set to zero). This is due to the fact that diffusion is an irreversible process.

For the applications in cosmology, the case when the pressure $p$ and the rest-frame energy density $\rho$ are connected by a linear relation $p= (\gamma-1)\rho$, $\gamma=const.$, is particularly important. By~\eqref{enthalpy} this can be achieved by choosing 
\begin{equation}\label{los}
h(n,s)=n^\gamma F(s),
\end{equation}
where $\gamma>0$ to ensure $\mathfrak{e}>0$ and $F'(s)>0$ to ensure positivity of $T$. In this situation, the entropy variable $s$ can be replaced by the modified entropy $S=F(s)$ and~\eqref{enteq} simplifies to
\begin{equation}\label{enteq2}
u^\mu\nabla_\mu S=\sigma n^{1-\gamma}.
\end{equation}

The consistent description of any matter dynamics in general relativity requires the coupling with the Einstein equation:
\begin{equation}\label{einstein}
R_{\mu\nu}-\frac{1}{2}g_{\mu\nu}R=T_{\mu\nu}\quad (8\pi G=c=1),
\end{equation} 
where $R_{\mu\nu}$ is the Ricci tensor of the metric $g$ and $R=g^{\mu\nu}R_{\mu\nu}$. 
However it is clear that fluid matter undergoing velocity diffusion cannot appear as the only source in the Einstein equation. In fact the Bianchi identity $\nabla^\mu(R_{\mu\nu}-\frac{1}{2}g_{\mu\nu}R)=0$ implies that solutions of the Einstein equation
can only exist if $\nabla_\mu T^{\mu\nu}=0$. 
The physical reason for this incompatibility is that our new fluid model does not take into account the interaction between the particles undergoing diffusion and the background medium in which diffusion takes place. 
While the diffusion approximation in Newtonian mechanics consists exactly in neglecting this interaction~\cite{calogero}, we see that in general relativity such an approximation, which leads inevitably to a model that does not preserve energy, is inconsistent. 
The next best ``diffusion approximation" consists in assuming the simplest possible interaction with the background medium. An interesting choice, because of its simplicity and because it leads to a natural generalization of the cosmological constant assumption, is to model the effect of the background medium by a scalar field  according to the following modification of~\eqref{einstein}:
\begin{equation}\label{einsteinphi}
R_{\mu\nu}-\frac{1}{2}g_{\mu\nu}R+\phi g_{\mu\nu}=T_{\mu\nu}\quad (8\pi G=c=1).
\end{equation} 
The scalar field $\phi$ will be called {\it cosmological scalar field}, since it enters in~\eqref{einsteinphi} in the place where the cosmological constant $\Lambda$ is commonly introduced. It should be emphasized once again that the cosmological scalar field is not to be interpreted as ordinary matter, but rather as a background field which, interacting with the fluid particles, causes their diffusion. According to particle physics, $\phi$ plays the role of  the vacuum energy. 

By~\eqref{divT}, the cosmological scalar field satisfies the equation
\begin{equation}\label{phieq}
\nabla_\mu\phi=\sigma n u_\mu.
\end{equation}
There are a number of important properties that one can infer from~\eqref{phieq}. Firstly,~\eqref{phieq} forces the matter flow to be irrotational:
\[\omega_{\mu\nu}=\nabla_\mu (nu_\nu)-\nabla_\nu (nu_\mu)=0,
\]
where $\omega_{\mu\nu}$ is the vorticity tensor of the fluid.
Secondly, $\phi$ satisfies the transport equation
\[
u^\mu\nabla_\mu\phi=-\sigma n,
\]
which shows that $\phi$ is decreasing along the matter flow. Finally, since the fluid number density is assumed to be a conserved quantity, i.e., $\nabla_\mu (n u^\mu)=0$,  the cosmological scalar field satisfies the homogeneous wave equation
\begin{equation}\label{waveeq}
\nabla^\mu\nabla_\mu\phi=0.
\end{equation}
It follows by~\eqref{waveeq} that the canonical energy-momentum tensor of the scalar field is divergence-free:
\[
\nabla_\mu S^{\mu\nu}=0,\ S_{\mu\nu}=\nabla_\mu\phi\nabla_\nu\phi-\frac{1}{2}g_{\mu\nu}\nabla^\alpha\phi\nabla_\alpha\phi.
\]
Note however that this energy-momentum tensor is not added as a source in the Einstein equation~\eqref{einsteinphi}. This again is due to our interpretation of $\phi$ as a background medium, rather than an ordinary matter field.

In conclusion the cosmological scalar field $\phi$ propagates through spacetime in form of waves without dissipation and interacts with the fluid particles causing their diffusion; as a consequence of this interaction, the scalar field $\phi$ is decreasing along the matter flow, which can be interpreted as vacuum energy being transferred to the fluid particles.  
In the absence of diffusion, i.e., when $\sigma=0$, $\phi$ is constant and~\eqref{einsteinphi} reduces to the Einstein equation with cosmological constant. 

\section{Cosmological solutions}
Next we consider spatially homogeneous and isotropic solutions, for which the line element can be written in the form
\begin{equation}\label{RW}
ds^2=-dt^2+a(t)^2\left[\frac{dr^2}{1-k r^2}+r^2d\Omega^2\right],
\end{equation}
where $a(t)>0$ is the scale factor, $k=0$ or $\pm 1$ is the curvature parameter of the hypersurfaces $t=const.$ and $d\Omega$ is the invariant surface element on the unit sphere. This class of solutions is the most important and widely used for the applications in cosmology~\cite{wein}. 
Let us write the fluid equations derived previously on a Lorentzian manifold with line element~\eqref{RW}. In this case the matter field variables are functions of $t$ only. Moreover we assume that $\rho=h(n,s)$ is given by~\eqref{los}, which implies that $p=(\gamma-1)\rho$, for some constant $\gamma$. Although this is not always necessary in what follows, it will be assumed that $2/3\leq\gamma<2$;
in particular the fluid satisfies the strong and the dominant energy condition. The conservation of particles number, eq.~\eqref{divJ}, implies
\begin{equation}\label{nRW}
n(t)=\frac{a_0^3n_0}{a(t)^3},
\end{equation}
where a subscript $0$ means evaluation at time $t=0$. The equation~\eqref{enteq2} on the modified entropy $S$ becomes
\begin{equation}\label{SRW}
\dot{S}=\sigma n_0^{1-\gamma}\left(\frac{a_0}{a(t)}\right)^{3-3\gamma},
\end{equation}
where a dot denotes differentiation with respect to $t$. Next the equations on the geometric fields will be considered.
The equation~\eqref{phieq} on the cosmological scalar field $\phi$ becomes
\begin{equation}\label{phiRW}
\dot{\phi}=-\sigma\frac{n_0a_0^3}{a(t)^3},
\end{equation}
while the Einstein equation~\eqref{einsteinphi} entails
\begin{align}
&H^2=\frac{1}{3}(\rho+\phi)-\frac{k}{a(t)^2},\label{hubble}\\
&\dot{H}=\frac{1}{3}\left[\phi-\left(\frac{3}{2}\gamma-1\right)\rho\right]-H^2,\label{Heq}
\end{align}
where 
\begin{equation}\label{rho}
\rho=\left(\frac{a_0^3n_0}{a(t)^3}\right)^\gamma S,\quad H=\frac{\dot{a}}{a};
\end{equation}
$H$ denotes the Hubble function.
For any specific choice of $\gamma$ and $k=0,\pm1$, the equations~\eqref{nRW}--\eqref{Heq} form a complete system for the fluid variables $(n,S)$, the scale factor $a$, the Hubble function $H$ and the cosmological scalar field $\phi$. The initial data consist of $a_0>0$, $n_0,S_0,\phi_0,H_0$ such that~\eqref{hubble} is satisfied at time $t=0$. Identifying $t=0$ with the present value of the cosmological time, it is physically justified to assume $H_0>0$ --- since the universe is currently expanding --- and $\phi_0\geq 0$ --- since $\phi_0$ corresponds to the present observed value of the cosmological constant. It will also be assumed of course that $n_0,S_0>0$. 

In the following the global behavior of solutions to~\eqref{nRW}--\eqref{Heq} in the flat case ($k=0$)  will be analyzed. Let us start by recalling the global behavior of the analogous diffusion-free cosmological model ($\sigma=0$). In the absence of diffusion, one recovers the Einstein-perfect fluid system with non-negative cosmological constant  $\Lambda=\phi_0$. In this case it is well known that the model develops a singularity in finite time in the past, while in the future it is singularity-free and forever expanding. (In fact, one can even solve explicitly the equations in this case, see~\cite[Sec. 14.2]{exact}.) It will now be shown that when diffusion is switched on, the past and future qualitative behavior of the universe may be  completely different.

Since $k=0$, we can introduce cartesian coordinates $(x,y,z)$ such that 
\[
ds^2=-dt^2+a(t)^2(dx^2+dy^2+dz^2)
\]
and $a_0=1$.

Since $\phi_0\geq 0$ and $\phi$ is decreasing, the cosmological scalar field is always positive in the past, causing accelerated expansion of the universe, and could become negative at later times.  If this happens the universe is doomed to collapse into a singularity in finite time.  In fact if $\phi(\bar{t})<0$ at some time $\bar{t}$, then by~\eqref{Heq} we will have $\dot{H}<-|\phi(\bar{t})|/3-H^2$ for all $t>\bar{t}$ 
and therefore there exists a time $t_*$ such that $H(t)\to -\infty$ as $t\to t_*$, which implies $a(t)\to 0$ as $t\to t_*$.  On the other hand, if $\phi(t)$ never vanishes, then the cosmological model will be singularity free and forever expanding in the future. This is a trivial consequence of~\eqref{hubble}$_{k=0}$, which shows that $H$ never vanishes if $\phi>0$, and therefore, since $H_0>0$, it will remain positive for all times. Let $\phi_\infty=\lim_{t\to\infty}\phi\geq 0$. It will now be shown that when $\phi_\infty>0$, the metric behaves asymptotically as the De Sitter solution, precisely
\begin{equation}\label{claim}
a(t)\sim \exp\left(\sqrt{\frac{\phi_\infty}{3}}t\right), \ \text{as }\ t\to\infty.
\end{equation}
In fact, by L'H\^opital's rule,
\[
\rho\sim\frac{S}{a^{3\gamma}}\sim\frac{\dot{S}}{a^{3\gamma-1}\dot{a}}\sim\frac{1}{a^3H}.
\]
Moreover by~\eqref{hubble}-\eqref{Heq},
\[
\frac{d}{dt}(a^3H)=\frac{3}{2}a^3H^2(2-\gamma)+\frac{\gamma}{2}a^3\phi\geq \frac{\gamma}{2}\phi_\infty.
\]
Whence $a^3H\to\infty$, and thus $\rho\to 0$, as $t\to\infty$. By by~\eqref{hubble}, $H^2\to\phi_\infty/3$, which implies~\eqref{claim}.
\begin{figure}[Ht]
\begin{center}
\subfigure[Future collapse ($S_0=3, \phi_0=0.3$)]{%
\includegraphics[width=0.5\textwidth]{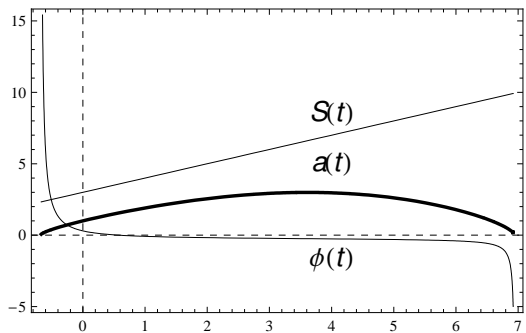}}
\qquad
\subfigure[Unlimited expansion ($S_0=3, \phi_0=1$)]{
\includegraphics[width=0.5\textwidth]{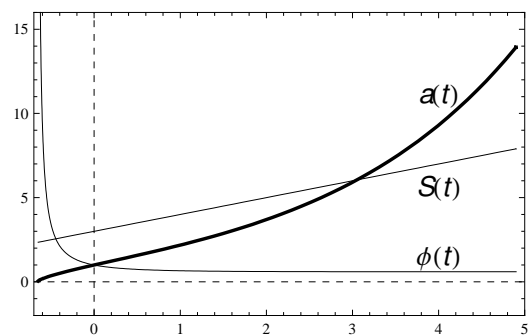}}
\caption{Examples of solutions with big-bang singularity ($\sigma=\gamma=n_0=a_0=1$)}
\label{bigbang}
\end{center}
\end{figure}

The behavior of the cosmological model in the past is more sensitive to the choice of the initial value $S_0$ rather than $\phi_0$. If $S_0$ is sufficiently small, the modified entropy $S$ could vanish at some negative time $t_0$, while $a(t_0)$ is still positive. (For instance, in the case of dust ($\gamma=1$), the solution of~\eqref{SRW} is $S(t)=S_0+\sigma t$ and therefore $t_0=-S_0/\sigma$.) If this happens, the energy density $\rho$ becomes negative and therefore the solution is unphysical for $t<t_0$, even if the metric of spacetime remains smooth. However a physically viable and interesting cosmological model can be constructed by matching the metric at the time $t_0$ with the De Sitter solution $a_{DS}(t)=C\exp (\sqrt{\phi(t_0)/3}\,t)$,
where the constant $C$ is such that $a_{DS}(t_0)=a(t_0)$.
The resulting cosmological model has no big-bang singularity and is vacuum up to the time $t_0$ (since $\rho=0$ and $a(t)=a_{DS}(t)$ for $t\leq t_0$), at which time the vacuum energy $\phi(t_0)$ starts to be converted into matter energy $\rho$. 
The other possibility is that the scalar factor vanishes while the modified entropy is still positive. In this case a big-bang singularity forms in the past.

\begin{figure}[Ht]
\begin{center}
\subfigure[Future collapse $(S_0=0.6,\phi_0=0.1)$]{%
\includegraphics[width=0.5\textwidth]{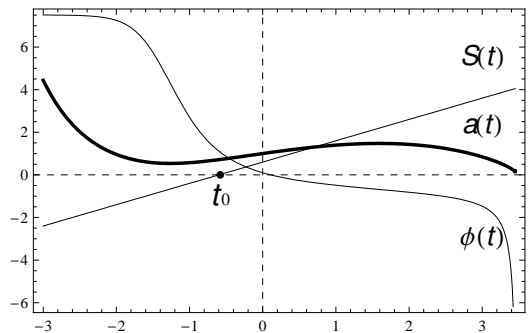}}
\qquad
\subfigure[Unlimited expansion $(S_0=0.6,\phi_0=1)$]{
\includegraphics[width=0.5\textwidth]{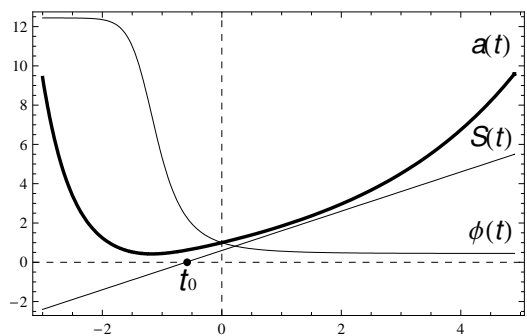}}
\caption{Examples of solutions without big-bang singularity ($\sigma=\gamma=n_0=a_0=1$). The solutions can be matched to De Sitter at the time $t=t_0=-0.6$. }
\label{nobigbang}
\end{center}
\end{figure}

Figures~\ref{bigbang} and~\ref{nobigbang} show examples of solutions obtained numerically. The solutions depicted in Figure 1 have a big-bang singularity in the past, while those on Figure 2 do not. Moreover, the solutions depicted are all dust solutions ($\gamma=1$) and the constants $\sigma,n_0$ have been fixed to one.  Thus $S(t)=S_0+t$ in all pictures. Similar solutions can be obtained for $k=\pm 1$ and for all other values of $\gamma\in [2/3,2)$ and $\sigma>0$ (no matter how small). Whether any of these solutions may represent the actual behavior of our universe requires of course experimental evidence. In particular, while observational constraints are available for the values of $(n_0, \rho_0, \phi_0, H_0)$, the possibility that $\sigma>0$ remains open.

We conclude by pointing out that a simple explicit solution of~\eqref{nRW}--\eqref{Heq} for $k=0$ and $\gamma=1$ is given by
\begin{subequations}\label{exact}
\begin{align}
&a(t)=1+\left(\frac{n_0\sigma}{2}\right)^{1/3}t,\\
&\phi(t)=\frac{(n_0\sigma/2)^{2/3}}{a(t)^2},\ S(t)=\left(\frac{2\sigma^2}{n_0}\right)^{1/3}a(t).
\end{align}
\end{subequations}
The only free parameter in the solution~\eqref{exact} is $n_0$, thus it is highly non-generic. The qualitative behavior of this solution is similar to the one depicted in Figure~\ref{bigbang}(b). Note also that for $\sigma=0$ the solution reduces to $(a,\phi,S)\equiv (1,0,0)$, which corresponds to Minkowski spacetime. Thus the solution~\eqref{exact} is a ``pure diffusion" one.  

\vspace{1cm}

\noindent {\bf Acknowledgments:} The author is most grateful to the anonymous referee for valuable comments on a previous version of the paper.


\begin{thebibliography}{1}

\bibitem{calogero}
S.~Calogero.
\newblock A kinetic theory of diffusion in general relativity with cosmological scalar field.
\newblock {\em J. Cosmo. Atrop. Phys.} 11/2011, 016.



\bibitem{frajan} J.~Franchi, Y.~Le~Jan.
\newblock Relativistic Diffusions and Schwarzschild Geometry.
\newblock {\em Comm.\ Pure\ Appl.\ Math.}, 60~:~187--251, 2007.


\bibitem{haba} Z.~Haba. 
\newblock Relativistic diffusion with friction on a pseudoriemannian manifold. 
\newblock {\em Class.\ Quant.\ Grav.}, 27~:~095021, 2010.

\bibitem{herr} J.~Hermann.
\newblock Diffusion in the general theory of relativity. 
\newblock {\em Phys.\ Rev.\ D},  82~:~024026, 2010.



\bibitem{exact}
H.~Stephani, D.~Kramer, M.~MacCallum, C.~Hoenselaers, E.~Herit.
\newblock {\em Exact solutions of Einstein's field equations}.
\newblock Cambridge Monographs on Mathematical Physics, Cambridge (UK), 2003.

 


\bibitem{wein}
S.~Weinberg.
\newblock {\em Cosmology.}
\newblock Oxford University Press, 2008.
\end{thebibliography}
\end{document}